\documentclass{aip-cp}

\usepackage[numbers]{natbib}
\usepackage{rotating}
\usepackage{graphicx}
\usepackage{wrapfig}

\begin{document}

\title{Radon Daughter Plate-out onto Teflon}
\author[aff1]{E.S. Morrison\corref{cor1}}
\author[aff1]{T. Frels}
\author[aff1]{E.H. Miller}
\author[aff1]{R.W. Schnee}
\author[aff1]{J. Street}

\affil[aff1]{Department of Physics, South Dakota School of Mines \& Technology, Rapid City, SD 57701 USA}
\corresp[cor1]{Eric.Morrison@mines.sdsmt.edu}
\maketitle

\begin{abstract}
Radiopure materials for detector components in rare event searches may be contaminated after manufacturing with long-lived $^{210}$Pb produced by the decay of atmospheric radon. Charged radon daughters deposited on the surface or implanted in the bulk of detector materials have the potential to cause noticeable backgrounds within dark matter regions of interest. Understanding the mechanics governing these background signals is therefore a paramount concern in dark matter experiments in order to distinguish a real signal from internal detector backgrounds. Teflon ($i.e.$ PTFE) is a specific material of interest because it makes up the walls of the inner detector of many liquid noble detectors such as the LUX-ZEPLIN experiment. The rate of radon daughter plate-out onto Teflon can be orders of magnitude larger than the plate-out rate onto other materials. Mitigation of plate-out onto Teflon and steel by proximity to other materials is demonstrated. 
\end{abstract}

\section{RADON PROGENY PLATE-OUT BACKGROUNDS}
\begin{figure}[t]
	\begin{tabular}{c c}
		\centering
  		\includegraphics[width=12cm, height=5cm]{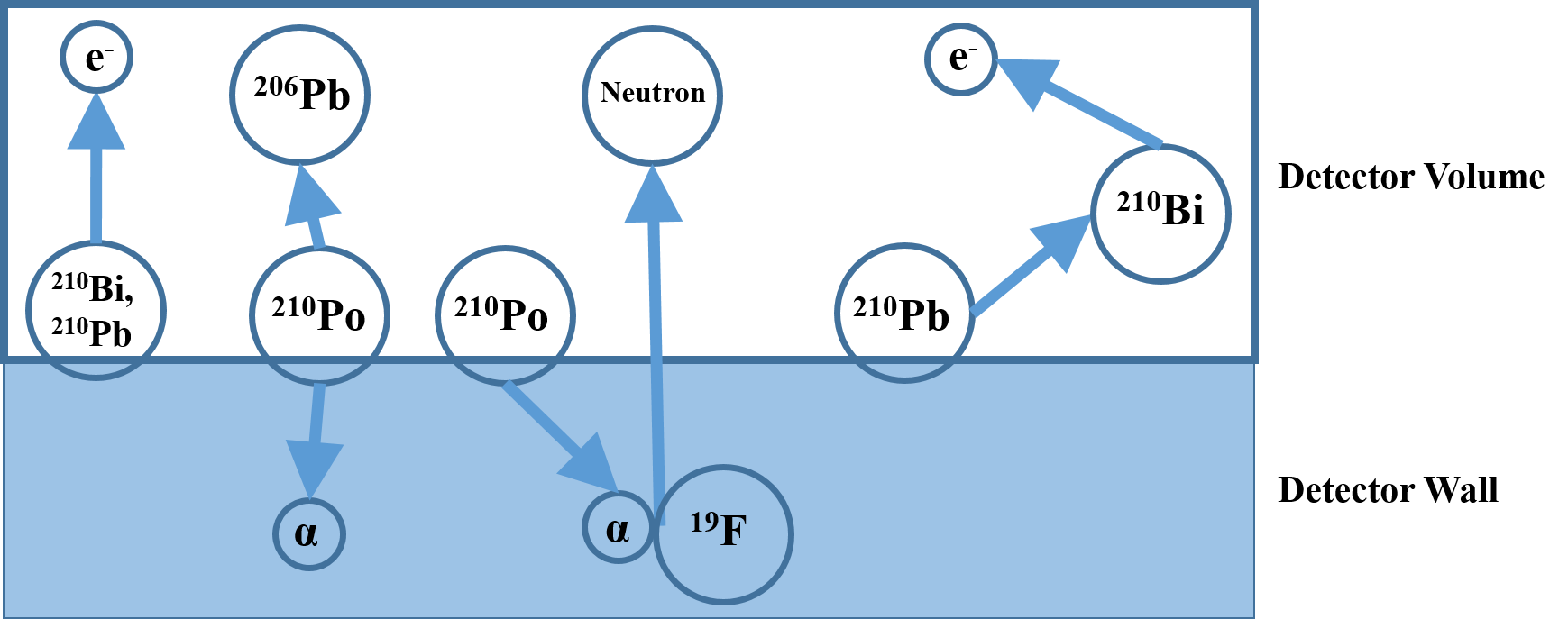} &
  		\includegraphics[width=3cm, height=5cm]{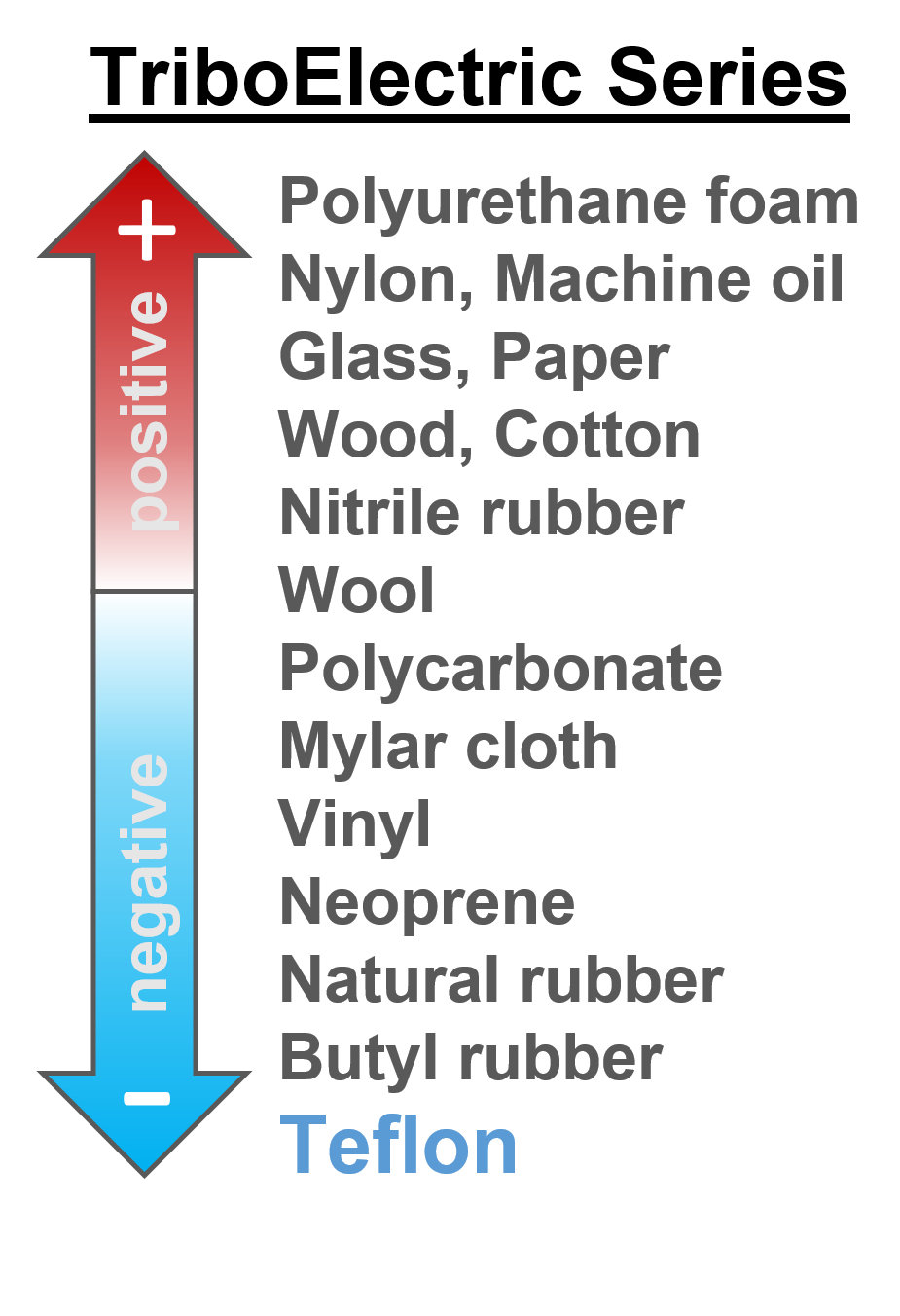}
  	\end{tabular}
  	\caption{\textit{Left}: Decay reactions from radon daughter plate-out on the surface of a detector wall that may produce prominent backgrounds in dark matter detectors. (Leftmost) A beta decay reaction of $^{210}$Bi or $^{210}$Pb relinquishes an electron into the volume generating an electron recoil signal. (2nd from the left) $^{210}$Po decay emits a $^{206}$Pb nucleus that recoils into the detector volume. (2nd from right) An alpha particle from $^{210}$Po decay on the surface may interact with a fluorine atom in the detector wall causing neutron emission in the volume. (Rightmost) A beta decay of $^{210}$Pb on the surface releases a $^{210}$Bi nucleus into the detector volume which itself may beta decay in the volume. \textit{Right}: Illustration of the positions of various materials in the triboelectric series. Teflon is on the far negative end of this spectrum, indicating that this material tends to carry a negative static charge~\cite{AlphaLab}.}
  	\label{BT}
\end{figure}

A principal contamination concern for dark matter experiments is radon progeny plate-out, a physical process in which charged radon daughter nuclei are deposited onto material surfaces from surrounding radon-laden air~\cite{LRT2004Leung,Mount:2017qzi}. Figure~\ref{BT} shows the cardinal radon progeny decay reactions responsible for backgrounds in dark matter detectors. These processes pose a challenge to dark matter detection in part because they deposit energy within the signal energy regions of interest. Decay of $^{210}$Pb on the surface of a detector wall may generate low-energy betas and x-rays that result in a near-surface electron-recoil background~\cite{PhysRevD.95.082002}. $^{210}$Bi undergoes $\beta$-decay with an $\sim$1.2\,MeV endpoint that also creates a near-surface electron-recoil background. Decay of$^{210}$Po emits an energetic alpha and a $^{206}$Pb nucleus with 103\,keV energy, potentially resulting in a continuum of nuclear-recoil signals up to $\sim$100\,keV depending on the implantation depth of the parent $^{210}$Po when the alpha is absorbed in inactive material. This recoiling nucleus causes  backgrounds in SuperCDMS~\cite{PhysRevD.95.082002} and LZ due to spatial leakage of poorly reconstructed events at the inner detector walls~\cite{Mount:2017qzi}. The alpha particle resulting from $^{210}$Po decay feeds ($\alpha,n$) reactions in detector materials such as Teflon~\cite{Mount:2017qzi}. Thus when an alpha particle interacts with a detector wall it may cause neutron emission into the detector volume. Decay of $^{210}$Pb on the surfaces of liquid detectors may release the daughter $^{210}$Bi nucleus into the liquid volume, where it may provide a ``naked'' beta decay~\cite{Mount:2017qzi}.

Understanding the origin of these backgrounds for Teflon is particularly important. The total surface area of Teflon used in the LZ experiment is 84\,m$^{2}$ and because it composes the walls of the inner detector, Teflon has the most surface area exposed to the active liquid xenon volume. Teflon has a large ($\alpha,n$) yield of 9.48 $\times$ 10$^{-6}$ neutrons/$\alpha$ due to its high fluorine content, making it a primary source of neutron emission backgrounds~\cite{Mount:2017qzi}. 

Plate-out rates onto materials are often estimated using the Jacobi model~\cite{jacobi1972,knutsonJacobi}, which simplifies the plate-out process by averaging over all atoms in the room rather than following individual atoms. For deposition within a cleanroom, the Jacobi model may be modified to ignore attachment of radon daughters to dust and consider the effective removal of radon daughters by filtration upon recirculation through cleanroom hepafilters. Deposition velocities $v$ are typically 5--15\,m/h~\cite{knutsonJacobi}. For a room of volume $V$,  surface area $S$, and air recirculation rate $R$, the deposition rate $\lambda_{\mathrm{D}} = vS/V$ and the radon daughter filtration rate $\lambda_{\mathrm{F}} =R/V$. For cleanrooms with high enough recirculation, essentially all plate-out occurs as $^{218}$Po because any $^{218}$Po that does not plate-out is filtered out before it decays. In equilibrium, the relationship between the radon concentration in the air $C_{\mathrm{air}}$ (atoms per unit volume) and the $^{218}$Po  activity on a surface $C_{\mathrm{S}}$ (atoms per unit area) is given by
\begin{equation}
\lambda_{1} C_{\mathrm{S}} = \lambda_{0} C_{\mathrm{air}} \times 
\frac{v}{ \lambda_{\mathrm{D}} + \lambda_{\mathrm{F} } + \lambda_{\mathrm{1}} }\equiv  \lambda_{0} C_{\mathrm{air}} \times h ,
\end{equation}
where $\lambda_0 = 0.00755$\,h$^{-1}$ is the $^{222}$Rn decay rate, $\lambda_1 = 13.37$\,hr$^{-1}$ is the $^{218}$Po decay rate, and the effective plate-out height $h$ is the height containing the number of radon atoms whose daughters will plate out onto a surface.  For typical cleanrooms, $h \sim 15$\,cm.

Based on such estimates, the LZ experiment imposes an exposure limit for Teflon components between 45--783\,days in its low-radon cleanroom~\cite{Mount:2017qzi}. However, Teflon occupies the far negative end of the triboelectric series (shown in Figure~\ref{BT}, \textit{Right}) while radon progeny nuclei have an 88\% probability to be positively charged~\cite{doi:10.1093/oxfordjournals.rpd.a006512}. Thus the plate-out rate onto Teflon is likely significantly higher than that onto other materials. Understanding the variation of radon daughter plate-out, particularly for Teflon, in a realistic setting is essential for planning detector assembly.

\section{EXPERIMENT PROCEDURE}
The rates of radon daughter plate-out onto Teflon and stainless steel samples were measured for 16 trials within a cleanroom with naturally high radon concentration ($\sim$100\,Bq/m$^{3}$). In each trial, samples collected plate-out for 2.5\,hours. Since the mean time of decay of the fast radon progeny through $^{214}$Po totals 50\,minutes, this collection time allowed the concentration of each daughter through $^{214}$Po to reach stability. The samples were then transferred to an Ortec AlphaDuo counter with a delay of only 1--2 minutes before counting began. The AlphaDuo measured $^{218}$Po and $^{214}$Po events from each sample for two hours, long enough for approximately all of the $^{214}$Po decays to occur. Measurements of radon activity in the air near the samples were recorded with a Durridge RAD7 radon detector, with interpolations (with higher uncertainties) required for times when the RAD7 pump failed.

\begin{figure}[!t]
	\begin{tabular}{c c}
		\centering
  		\includegraphics[width=10cm, height=7cm]{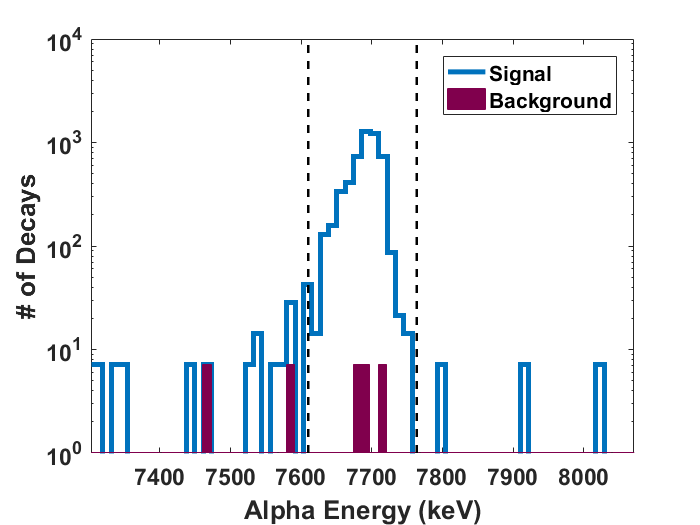}&
		\includegraphics[width=6cm, height=7cm]{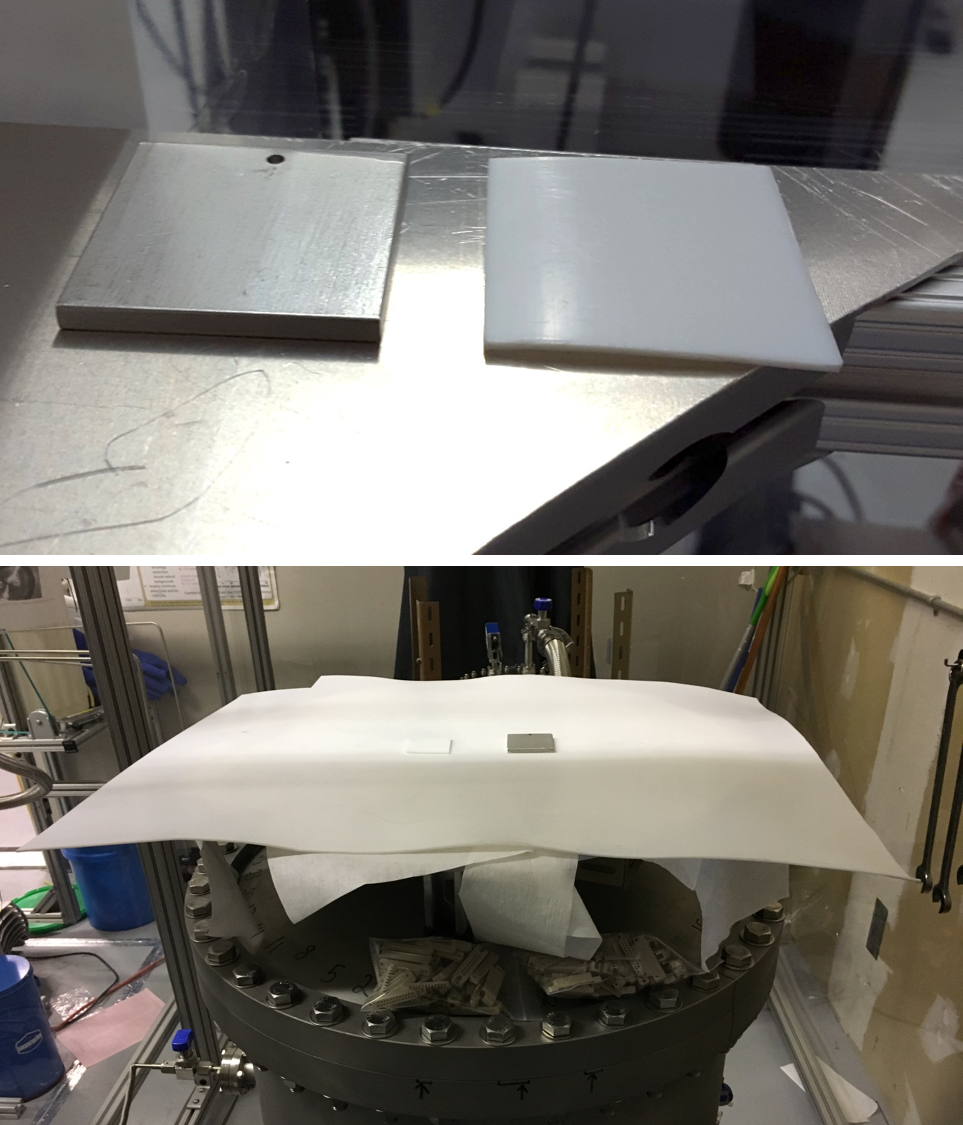}
  	\end{tabular}
  	\caption{\textit{Left}: Energy spectrum around the $^{214}$Po decay region of interest (outlined by black dashed lines) in detector 2 of the Ortec AlphaDuo counter, comparing the  efficiency-corrected signal from the first Teflon sample trial (solid unfilled blue lines) to the background expected (filled red) based on five 		background runs. \textit{Right top}: 2'' $\times$ 2'' square Teflon and steel samples on top of an aluminum shelf near acrylic walls. \textit{Right bottom}: Teflon and steel samples on the surface of a large Teflon sheet in the center of a cleanroom.}
  	\label{BSS}
\end{figure}

Runs of the Ortec AlphaDuo counter with no samples were completed in order to quantify background $^{214}$Po decays on the detector. Five background runs, each lasting two hours to mimic actual sample counting time, yielded only three counts total in the energy region of interest.
Figure~\ref{BSS}  shows the average background spectrum compared to that of the the first Teflon trial. No background subtraction was employed in the analysis described below.
 
In equilibrium, the rate at which $^{218}$Po atoms increase on a sample of area $A$ due to plate-out is the same as each daughter's decay rate:
\begin{equation}
 \frac{dC_{\mathrm{air}}}{dt} hA = \frac{dN_{i}}{dt} =  \frac{N_{i}}{\tau_i}  \end{equation}
 where $\tau_i$ is the lifetime and $N_i$ is the number of atoms on the sample of each progeny $i$ for the  progeny $^{218}$Po, $^{214}$Pb, $^{214}$Bi, and $^{214}$Po.  Because the counting time is long enough that essentially all radon progeny decay through $^{214}$Po, the total number of $^{214}$Po decays is the same as the total number of atoms that have plated out, 
\begin{equation}
\sum{N_i} = \sum{\tau_i} \frac{dC_{\mathrm{air}}}{dt} hA.
\end{equation} 
The number of detected $^{214}$Po decays $N_{\mathrm{events}} = \epsilon \sum{N_i} $, where $\epsilon= 13.85\%$ is the detection efficiency, and $dC_{\mathrm{air}}/dt = \lambda_{0} C_{\mathrm{air}}$. 
Solving for $h$ yields
 \begin{equation}
 h = \frac{N_{\mathrm{events}}} { \lambda_{0} C_{\mathrm{air}} A \epsilon \Sigma \tau_i}.
\end{equation}

As shown in Figure~\ref{BSS}, square 2'' $\times$ 2'' steel and Teflon samples were placed at specific locations within a Class-2000 cleanroom.  The first five trials involved exposing steel and Teflon samples on an aluminum sheet near the acrylic wall of the cleanroom. The samples and aluminum sheet were then moved to the center of the cleanroom for two more trials. For these first seven trials the air recirculation rate $R=750$\,cfm. The next four trials also occurred in the center of the cleanroom but the aluminum sheet was replaced by a sheet of plastic wrap in order to test the effectiveness of plastic wrap on diverting radon daughter plate-out. In the final five trials plastic wrap was substituted for a large Teflon sheet (5244\,cm$^{2}$ surface area) in order to test the hypothesis that a Teflon sheet will reduce plate-out onto smaller samples because the plate-out volume in the air is evenly distributed over the surface area of the sheet.  For the final nine trials the air recirculation rate $R=400$\,cfm. 

\begin{figure}[!t]
	\begin{tabular}{c c}
		\centering
  		\includegraphics[height=7.5cm]{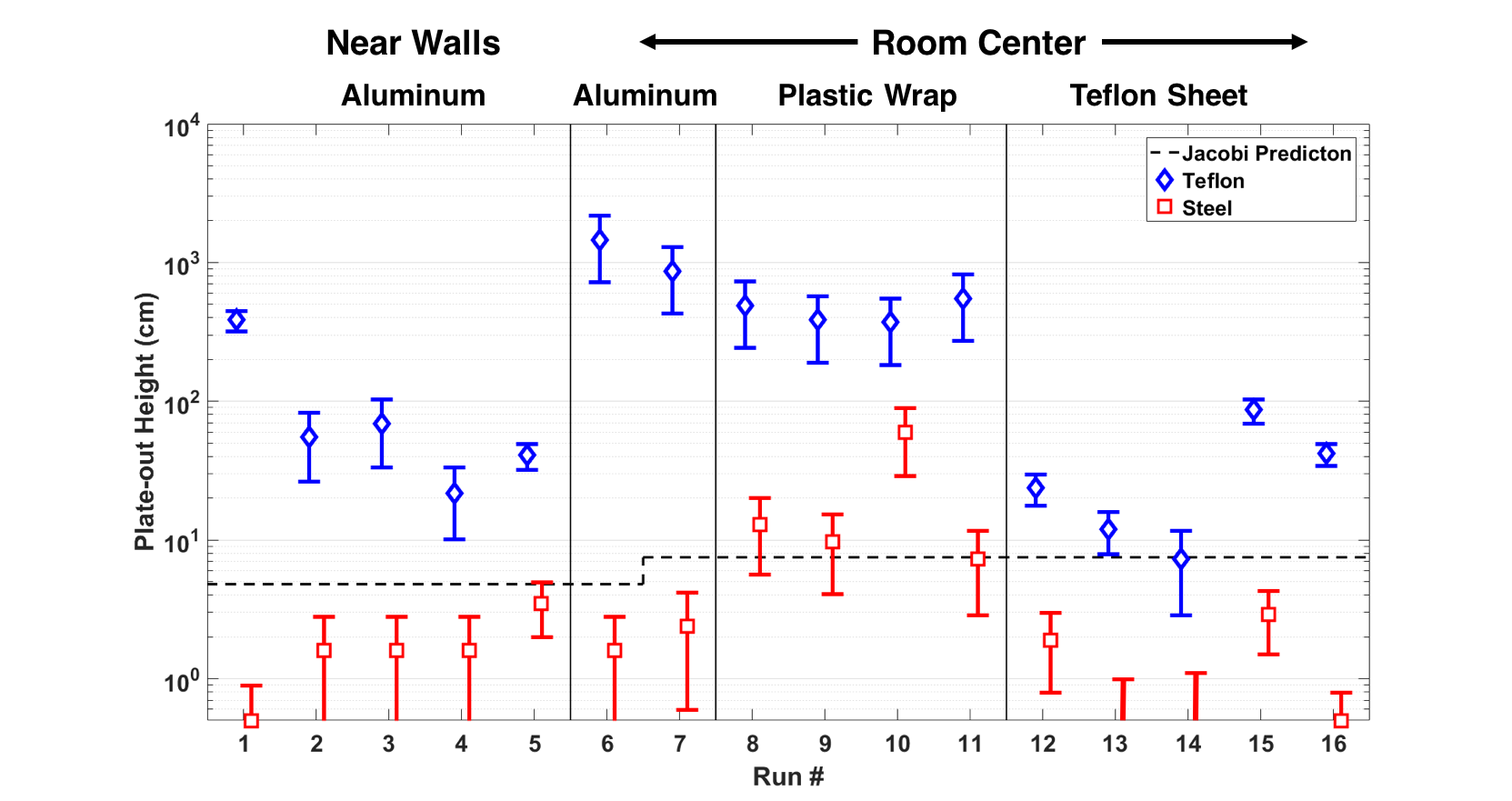}
  	\end{tabular}
  	\caption{Inferred plate-out height  for each run of the experiment for both Teflon (blue diamonds) and steel samples (red squares).  For most runs, the plate-out height onto Teflon is higher, and that onto steel is lower, than that predicted by the modified Jacobi model in the text (dashed line).  Headers indicate the positions of the samples within the room and the material on which the samples were placed.}
  	\label{Data}
\end{figure}

\section{EVALUATION OF RESULTS AND FUTURE WORK}

Figure~\ref{Data} shows inferred plate-out heights based on the data for all trials of the experiment, compared to the plate-out height expected from the modified Jacobi model.  The cleanroom had volume $V = 7.9$\,m$^{3}$ and surface area $S = 27.2$\,m$^{2}$, and so had an initial (final) filtration rate 
$\lambda_{\mathrm{F}} = 161$\,h$^{-1}$ ($\lambda_{\mathrm{F}} = 86$\,h$^{-1}$) and  
an expected deposition rate  $\lambda_{\mathrm{D}} \approx 34$\,h$^{-1}$, for an initial (final) expected effective plate-out height $h \approx 4.8$\,cm ($h \approx 7.5$\,cm). 
The first five trials near the acrylic wall yielded a  plate-out height for Teflon 
that averaged 40$\times$ higher than that for steel. The inferred Teflon plate-out height varied more than the steel plate-out height did. Teflon samples on aluminum located in the center of the room had an average plate-out height $\sim$10$\times$ greater than Teflon samples on aluminum near the acrylic wall. This change is about that to be expected from moving farther from the walls, the room's hepafilter, and the ceiling.  Steel samples on an aluminum sheet did not undergo a change in plate-out height when moved from near the acrylic wall to the center of the cleanroom. Replacing the aluminum sheet with plastic wrap reduced the effective plate-out height onto Teflon by $\sim2\times$ but increased the average plate-out height onto steel by $\sim$8$\times$. The large Teflon sheet successfully diverted radon daughter nuclei from small Teflon and steel samples on its surface, confirming our expectation based on Teflon's position in the triboelectric series. Teflon appears to attract radon progeny at a rate $\sim50\times$ higher than steel from these results, but there is likely a large variation based on geometry, air flow, and handling. The use of Teflon sheets to attract radon daughters away from sensitive materials during assembly may be an effective plate-out mitigation technique. Future tests will involve moving the large Teflon sheet around the cleanroom to find which location affects samples the greatest. Other methods to reduce radon plate-out, such as using a high voltage wire, will be tested as well.

\section{ACKNOWLEDGMENTS}
This work was supported in part by the National Science Foundation (Grant No. PHY-1506033) and the 
 Department of Energy (Grants No. DE-SC0014223 and DE-AC02-05CH11231).
 
\bibliography{RadonPlateoutFinal}
\bibliographystyle{plain}

\end{document}